\documentclass[onecolumn, a4size, draftcls, 12pt]{IEEEtran}
\usepackage{amsmath, amssymb, amsfonts}
\usepackage{graphicx}
\usepackage{epsfig}
\usepackage{caption}
\usepackage{cases}
\usepackage{tensor}
\usepackage{fancyhdr}
\usepackage{setspace}

\usepackage{dblfloatfix}
\usepackage{subcaption}

\usepackage{epstopdf}

\title{Improper Signaling for Symbol Error Rate Minimization in $K$-User Interference Channel
\footnote{This work has been submitted in part to the \emph{IEEE Global Communications Conference (GLOBECOM)}, Austin, Texas, USA, Apr. 2014.} 
\footnote{H. D. Nguyen and S. Sun are with the Institute for Infocomm Research (I2R), the Agency for Science, Technology and Research (ASTAR), Singapore (email(s): \{nguyendh, sunsm\}@i2r.a-star.edu.sg). }
\footnote{R. Zhang is with the Department of Electrical and Computer Engineering, National University of Singapore (email: elezhang@nus.edu.sg). He is also with the I2R, ASTAR, Singapore.}}

\author{\vspace{0.0in} Hieu Duy Nguyen, Rui Zhang, and Sumei Sun}

\begin{document}
\maketitle

\setlength{\baselineskip}{1.1\baselineskip}
\newtheorem{definition}{\underline{Definition}}[section]
\newtheorem{fact}{Fact}
\newtheorem{assumption}{\underline{Assumption}}[section]
\newtheorem{theorem}{\underline{Theorem}}[section]
\newtheorem{lemma}{\underline{Lemma}}[section]
\newtheorem{corollary}{\underline{Corollary}}[section]
\newtheorem{proposition}{\underline{Proposition}}[section]
\newtheorem{example}{\underline{Example}}[section]
\newtheorem{remark}{\underline{Remark}}[section]
\newtheorem{algorithm}{\underline{Algorithm}}[section]
\newcommand{\mv}[1]{\mbox{\boldmath{$ #1 $}}}

\captionsetup[table]{labelsep=newline}

\begin{abstract}
The rate maximization for the $K$-user interference channels (ICs) has been investigated extensively in the literature. However, the dual problem of minimizing the error probability with given signal modulations and/or data rates of the users is less exploited. In this paper, by utilizing the additional degrees of freedom attained from the \emph{improper signaling} (versus the conventional proper signaling), we optimize the precoding matrices for the $K$-user single-input single-output (SISO) ICs to achieve minimal pair-wise error probability (PEP) and symbol error rate (SER) with two proposed algorithms, respectively. Compared to conventional proper signaling as well as other state-of-the-art improper signaling designs, our proposed improper signaling schemes achieve notable SER improvement in SISO-ICs under both additive white Gaussian noise (AWGN) channel and cellular system setups. Our study provides another viewpoint for optimizing transmissions in ICs and further justifies the practical benefit of improper signaling in interference-limited communication systems.   
\end{abstract}

\section{Introduction}\label{sec:intro}

In next generation cellular systems, the high data rate demand requires a more efficient utilization of the limited available spectrum. The universal frequency reuse thus becomes more favourable, which however leads to more severe interference issues as compared to the traditional case with only a fractional frequency reuse. Interference is therefore a dominant limiting factor for the performance of future wireless communication networks. 

Interference channel (IC) is a fundamental model for multiuser wireless communication and has been extensively studied to date. A complete characterization of the capacity region of the IC corrupted by additive Gaussian noise, however, is still open, even for the simplest two-user case \cite{Etkin01}. In the finite-SNR regime, significant contributions have been made to the problem of characterizing the rate region of ICs. For example, coordinated precoding/beamforming has been proposed to be implemented among BSs to control the inter-cell interference (ICI) to their best effort \cite{Dahrouj01}, \cite{Zhang01}, and various parametrical characterizations of the Pareto boundary of the achievable rate region have been obtained for the multiple-input single-output (MISO)-IC with coordinated transmit beamforming and single-user detection \cite{Zhang01}-\cite{Bjornson01}.

The recent advance in the so-called interference alignment (IA) technique has motivated numerous studies on characterizing the rate performance of ICs under the high signal-to-noise ratio (SNR) regime. With the aid of IA, the maximum achievable rates in terms of degree of freedom (DoF) have been obtained for various IC models to provide useful insights on designing optimal transmission schemes for interference-limited communication systems (see, e.g., \cite{Jafar01} and the references therein).

Another notable advancement is the use of improper Gaussian signaling (IGS) for ICs. Different from conventional systems employing proper Gaussian signals whose real and imaginary parts have equal power and are independent zero-mean Gaussian random variables, the real and imaginary parts of improper Gaussian signals have unequal power and/or are correlated \cite{Schreier01}. Improper signals have been investigated in applications such as detection and estimation \cite{Schreier01}-\cite{Navarro-Moreno01}. Studies of IGS in communication systems, however, only appeared recently. This may be due to the fact that proper Gaussian signaling (PGS) has been known to be capacity optimal for the Gaussian point-to-point, multiple-access, and broadcast channels; as a result, it was presumably deemed to be optimal for ICs. In \cite{Cadambe02}, it has been shown that IGS can further improve the achievable rates of the three-user IC in high-SNR regime. Inspired by this work, subsequent studies have investigated ICs with IGS in finite-SNR regime. Particularly, \cite{Ho01} and \cite{Zeng01} studied the achievable rate region of the two-user SISO IC. The rate region and minimum signal to interference-plus-noise ratio (SINR) maximization for the $K$-user IC have been characterized in \cite{Park02} and \cite{Zeng02}. These works have reported a significant rate improvement of IGS over conventional PGS in terms of achievable rate under finite SNR.

It comes to our attention that most of the existing work on ICs has focused on investigating the rate performance. However, a dual problem for ICs, which minimizes the transmission error probability with given users' signal modulations and/or data rates, is less exploited. Notice that this problem may be more practically sensible for the scenarios when the users have their desired quality-of-service (QoS) in terms of data rate and error performance to be met. It is worth noting that in \cite{Shen01} and \cite{Chen01}, the authors considered the problem of minimizing the mean squared error (MSE) in ICs to indirectly minimize the error rate. Although MSE is a meaningful criterion in practice, minimizing the MSEs in ICs does not necessarily lead to the error probability minimization. Moreover, \cite{Choi01} has studied the error performance for ICs based on IA. However, it is restricted only to the case of three-user ICs with at least two antennas at each node, in which each of the three user links can achieve at least one DoF. In contrast to the above prior work, in this paper, we study the problem of minimizing the users' symbol error rates (SERs) directly in the $K$-user single-input single-output (SISO) IC by applying improper signaling over finite signal constellations. We are motivated by the results that IGS can provide rate gains for ICs over the conventional PGS \cite{Cadambe02}-\cite{Zeng02}. It is thus expected that the additional degrees of freedom provided by improper signaling can also be exploited to improve the SER performance in ICs, even with practical (non-Gaussian) modulation schemes. 

Our study is also related to the classic problem of constellation design in digital communication. For additive white Gaussian noise (AWGN) channel, the design and analysis for various digital modulation schemes can be found in the early investigation (see, e.g., \cite{Cahn01} and \cite{Foschini01}). The error probability for fading channels was studied in, e.g., \cite{Beaulieu01}. The work \cite{Craig01} reported an important representation of the Q-function for Gaussian distribution, which has been widely used in subsequent works on the error rate analysis for digital modulation. More information on the constellation design for AWGN and fading channels can be found in the classic book \cite{Proadkis01}. With the introduction of multiple antennas, the constellation design for multiple-input multiple-output (MIMO) channels has attracted significant attention. Some notable results can be found in, e.g., \cite{Hochwald01} and \cite{Xin01}. However, there has been considerably less studies on the constellation design and error rate analysis for ICs.  

In this paper, we investigate the $K$-user IC with fixed signal modulations and data rates of the users. Different from the conventional setup where proper signaling is assumed, here we employ improper signaling to improve the error performance in ICs. We first derive the pair-wise error probability (PEP) of erroneously decoding one user's symbol to another, and formulate the precoding optimization problems to minimize the tranmission error probability according to two criteria, i.e., PEP and SER. Based on these results, two improper signalling schemes are proposed to minimize the maximum PEP and SER of all users, respectively. Numerical results show that an improved error rate performance can be achieved by the proposed schemes over conventional proper signaling as well as other state-of-the-art improper signaling designs, under both AWGN channel and practical cellular system setups.

The rest of the paper is organized as follows. Section \ref{sec:system model} describes the $K$-user SISO IC model and introduces improper signaling. In Sections \ref{sec:min max PEP} and \ref{sec:min SER upper-bound}, we present our new approaches to directly minimize the maximum PEP and SER of users in IC with two proposed algorithms, respectively. Section \ref{sec:benchmark} presents various benchmark schemes employing conventional proper signalling or other improper signaling designs for comparison. Numerical results and relevant discussions are given in Section \ref{sec:numerical results}. Finally, Section \ref{sec:conclusions} concludes the paper.

{\it Notations}: Scalars and vectors/matrices are denoted by lower-case and bold-face lower-case/upper-case letters, respectively. The conjugate, transpose, and conjugate transpose operators are denoted as $(\cdot)^*$, $(\cdot)^T$, and $(\cdot)^H$, respectively. $[\mv{A}]_{i,j}$ represents the $(i,j)$-th element of the matrix $\mv{A}$. $\mathbb{E}[\cdot]$ denotes the statistical expectation. $\mv{Tr}(\cdot)$ represents the trace of a matrix. The distribution of a circularly symmetric complex Gaussian (CSCG) and real Gaussian random variable (RV) with zero mean and covariance $\sigma^2$ are denoted by $\mathcal{CN}(0,\sigma^2)$ and $\mathcal{N}(0,\sigma^2)$, respectively; and $\sim$ stands for ``distributed as''. $\mathbb{C}^{x \times y}$ denotes the space of $x\times y$ complex matrices.

\section{System Model}\label{sec:system model}

\begin{figure}[t]
 		\centering 
 		\epsfxsize=0.48\linewidth
 		\captionsetup{width=0.65\textwidth} 
 		\includegraphics[width=10.5cm, height=8.5cm]{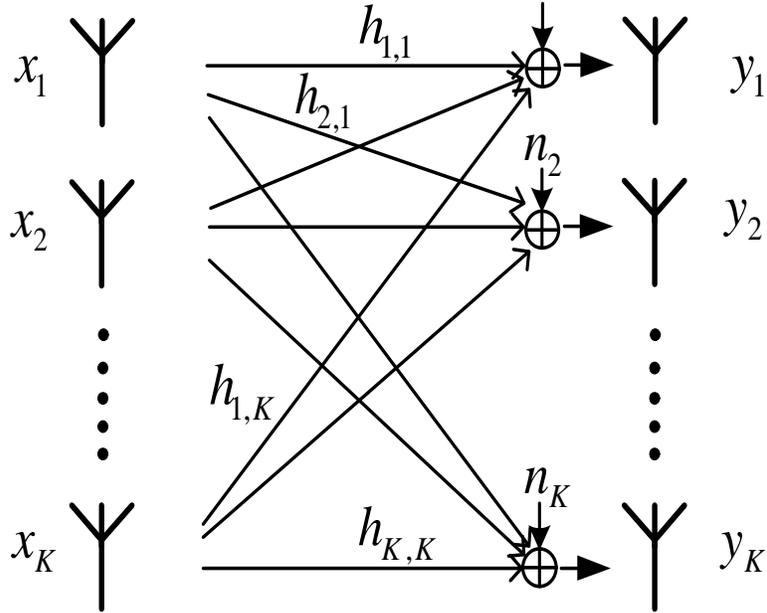}
 		\caption{The $K$-user SISO IC.}\label{fig:KIC}
\end{figure}

We consider a $K$-user SISO IC as shown in Fig. \ref{fig:KIC}, where the received complex baseband signals for the $k$-th user is expressed as
\begin{align}\label{eq:sys mod}
& y_k = h_{k1}x_1 + h_{k2}x_2 + \dots + h_{kK}x_K + n_k, 
\end{align}
where $h_{kl} = |h_{kl}|e^{j\theta_{kl}}$, $k$, $l$ $\in$ $\{1$, \dots, $K\}$, is the complex coefficient for the channel between transmitter $l$ and receiver $k$; $x_k$ is the transmitted symbol for user $k$; and $n_k$ is the AWGN at receiver $k$, which is assumed to be a CSCG RV, denoted by $n_k$ $\sim\mathcal{CN}(0,\sigma_k^2)$. The transmit power of user $k$ is assumed to be limited by $P_{k}$, i.e., $\mathbb{E}[||x_{k}||^2]$ $\leq$ $P_{k}$. For convenience, we define the SNR of user $k$ as $\text{SNR}_k$ = $P_{k}/\sigma_k^2$. In this study, we assume that each receiver employs the practical single-user detection in decoding the desired signal and hence treats the interference from all other users as additional noise. However, different from the conventional setup where proper signaling is assumed at transmitters, we consider the use of more general improper signals. We first define the propriety and impropriety for a complex RV as follows.

\begin{definition}[\cite{Schreier01}]
Assume that a zero-mean random vector $\mv{v}$ $\in$ $\mathbb{C}^{n\times 1}$ has the covariance and pseudo-covariance matrices defined as $\mv{C}_{v}$ $\triangleq$ $\mathbb{E}[\mv{v}\mv{v}^H]$ and $\widetilde{\mv{C}}_{v}$ $\triangleq$ $\mathbb{E}[\mv{v}\mv{v}^T]$, respectively. The random vector $\mv{v}$ is called proper if $\widetilde{\mv{C}}_{v}$ $=$ $\mv{0}$. Otherwise, it is called improper.
\end{definition}

A special case of the above definition with $n=1$ is stated in the following. 
\begin{definition}
Given a zero-mean complex RV $\alpha = \alpha_R + j\alpha_I$ and its real covariance matrix defined as $\mv{C}_{\alpha}$ = $\mathbb{E}\left[ [\alpha_R ~~ \alpha_I]^T [\alpha_R ~~ \alpha_I] \right]$. The RV $\alpha$ is called proper if $\mv{C}_{\alpha}$ is a scaled identity matrix, i.e., $\mv{C}_{\alpha}$ = $p\mv{I}$ with $p>0$, which means that the real and imaginary parts $\alpha_R$ and $\alpha_I$ are uncorrelated and have equal variance of $p$. Otherwise, we call $\alpha$ improper. 
\end{definition}

It is worth noting that in practical digital communication systems, modulation schemes such as PSK (e.g., QPSK, 8PSK) and square QAM (e.g., 16QAM, 64QAM) all have the signal constellations drawn from proper RVs. For the convenience of our analysis in the sequel of this paper, we use the following equivalent real-valued representation of the complex-valued system in (\ref{eq:sys mod}), which is essentially a $K$-user $2\times 2$ MIMO IC with all real entries given by
\begin{align}\label{eq:real-valued IC}
\underbrace{ \begin{bmatrix} y_{kR} \\ y_{kI} \end{bmatrix} }_{\triangleq \mv{y}_k}
= |h_{kk}| \underbrace{ \begin{bmatrix} \cos{\theta_{kk}} & -\sin{\theta_{kk}} \\ \sin{\theta_{kk}} & \cos{\theta_{kk}} \end{bmatrix} }_{\triangleq \mv{J}(\theta_{kk})} \underbrace{ \begin{bmatrix} x_{kR} \\ x_{kI} \end{bmatrix} }_{\triangleq \mv{x}_k}
+ \sum_{l=1, l\neq k}^K |h_{kl}|
\underbrace{ \begin{bmatrix} \cos{\theta_{kl}} & -\sin{\theta_{kl}} \\ \sin{\theta_{kl}} & \cos{\theta_{kl}} \end{bmatrix} }_{\triangleq \mv{J}(\theta_{kl})}
\underbrace{ \begin{bmatrix} x_{lR} \\ x_{lI} \end{bmatrix} }_{\triangleq \mv{x}_{l}}
+ \underbrace{ \begin{bmatrix} n_{kR} \\ n_{kI} \end{bmatrix} }_{\mv{n}_k}
\end{align}
for $k=1, \dots, K$, where ``$R$'' and ``$I$'' denote the real and imaginary parts, respectively; and $n_{kR}$ and $n_{kI}$ are independent and identically distributed (i.i.d.) real Gaussian RVs each distributed as $\mathcal{N}(0,\sigma_k^2/2)$. 

Consider a normalized constellation for each user $k$, represented by a pair of real symbols in $\mv{d}_k = [d_{kR} ~~ d_{kI}]^T$ with identity covariance matrix $\mathbb{E}\left[ \mv{d}_k \mv{d}_k^T \right]$ $=\mv{I}$. For example, the symbol set for a normalized constellation from QPSK is given by
\begin{align}
\left\{ \begin{bmatrix} 1 \\ 1\end{bmatrix}, \begin{bmatrix} 1 \\ -1\end{bmatrix}, \begin{bmatrix} -1 \\ -1\end{bmatrix}, \begin{bmatrix} -1 \\ 1\end{bmatrix} \right\}.
\end{align}

The transmitted symbol $\mv{x}_k$ with improper signalling can then be obtained with the following transformation:
\begin{align}\label{eq:transformation}
\begin{bmatrix} x_{kR} \\ x_{kI} \end{bmatrix}
= \underbrace{ \begin{bmatrix} a_{k,11} & a_{k,12} \\ a_{k,21} & a_{k,22} \end{bmatrix} }_{\triangleq \mv{A}_k} 
\begin{bmatrix} d_{kR} \\ d_{kI} \end{bmatrix}, 
\end{align}
or equivalently $\mv{x_k} = \mv{A}_k \mv{d_k}$. Here $\mv{A}_k$ is the precoding matrix, while the linear operation in (\ref{eq:transformation}) is also called the widely linear processing \cite{Schreier01}, \cite{Navarro-Moreno01}. Alternatively, $\mv{A}_k$ can be regarded as a rotation and scaling matrix applied over the proper signal constellation of user $k$ to obtain improper signals. Furthermore, the transmit power constraint for user $k$ is re-expressed as $\mv{Tr}\left\{\mv{A}_k \mv{A}_k^T\right\}$ $=$ $a_{k,11}^2 + a_{k,12}^2 + a_{k,21}^2 + a_{k,22}^2 \leq P_{k}$.

From (\ref{eq:real-valued IC}), the real system model of the $K$-user IC with improper signaling can be expressed in the following form,
\begin{align}
\mv{y}_k = |h_{kk}| \mv{J}(\theta_{kk}) \mv{A}_k\mv{d}_k + \sum_{l=1, l\neq k}^K |h_{kl}|\mv{J}(\theta_{kl}) \mv{A}_{l} \mv{d}_{l} + \mv{n}_k.
\end{align}

Without loss of generality, we assume that each user $k$ applies the decoding matrix at the receiver in the form of $\mv{J}(\theta_{kk})\mv{R}_k$. The signal after applying the decodng matrix is given by
\begin{align}\label{eq:output r_k}
\mv{r}_k = \mv{R}_k^T \mv{J}^T(\theta_{kk}) \mv{y}_k = |h_{kk}| \mv{R}_k^T \mv{A}_k\mv{d}_k + \sum_{l=1, l\neq k}^K |h_{kl}| \mv{R}_k^T \mv{J}(\phi_{kl}) \mv{A}_{l} \mv{d}_{l} + \mv{R}_k^T \tilde{\mv{n}}_k,
\end{align} 
where $\phi_{kl} = \theta_{kl} - \theta_{kk}$, $k, l \in \{1, \dots, K\}$ and $l\neq k$; and $\tilde{\mv{n}}_k$ $=$ $\mv{J}^T(\theta_{kk})\mv{n}_k$ $\triangleq$ $[\tilde{n}_{kR}$ $\tilde{n}_{kI}]^T$ with $\tilde{n}_{kR}$ and $\tilde{n}_{kI}$ being i.i.d. Gaussian RVs each distributed as $\mathcal{N}(0,\sigma_k^2/2)$, $k=1, \dots, K$. 


Denote $\mv{w}_k$ = $\sum_{l=1, l\neq k}^K$ ${|h_{kl}|}$ $\mv{J}(\phi_{kl})$ $\mv{A}_{l}$ $\mv{d}_{l}$ + $\mv{n}_{k}$ as the effective noise at the receiver of user $k$, which includes both additive noise and interference. Then the post-processed signal in (\ref{eq:output r_k}) can be re-expressed as
\begin{align}\label{eq:output r_k 2}
\mv{r}_k = |h_{kk}|\mv{R}_k^T\mv{A}_k\mv{d}_k + \mv{R}_k^T\mv{w}_k.
\end{align}

In this paper, we assume that the $k$-th receiver estimates $\mv{W}_k$, where $\mv{W}_k$ is defined in (\ref{eq:INR CovMat}) below, and applies the whitening filter to the effective noise as the decoding matrix, i.e., $\mv{R}_k$ $=$ $\mv{W}_k^{-1/2}$.
\begin{align}\label{eq:INR CovMat}
\mv{W}_k = \mathbb{E}\left[ \mv{w}_k\mv{w}_k^T \right] = \frac{\sigma_k^2}{2}\mv{I} + \sum_{l=1, l\neq k}^K |h_{kl}|^2 \mv{J}(\phi_{kl}) \mv{A}_{l} \mv{A}_{l}^T \mv{J}^T(\phi_{kl}).
\end{align}

Considering the general $M_k$-ary ($M_k>1$) modulation for $\mv{d}_k$ of each user $k$, the (exact) SER expression is difficult to be obtained even assuming the conventional proper signalling, i.e., $\mv{A}_k$ is a scaled identity matrix for all $k$'s. Therefore, in this paper, we approximate the SER with improper signalling by an upper bound, which is obtained by applying the union bound as follows:
\begin{align}\label{eq:SERs of 2 users}
\text{SER}_k \leq \frac{1}{M_k-1} \sum_{\mv{d}_k} \sum_{\mv{\tilde{d}}_k \neq \mv{d}_k} Pr\{\mv{d}_k \to \mv{\tilde{d}}_k|\mv{d}_k\} Pr\{\mv{d}_k\} = \frac{1}{M_k(M_k-1)} \sum_{\mv{d}_k} \sum_{\mv{\tilde{d}}_k \neq \mv{d}_k} Pr\{\mv{d}_k \to \mv{\tilde{d}}_k|\mv{d}_k\},
\end{align}
where $Pr\{\mv{d}_k \to \mv{\tilde{d}}_k|\mv{d}_k\}$ is the so-called PEP when $\mv{d}_k$ is erroneously decoded as $\mv{\tilde{d}}_k$ with $\mv{\tilde{d}}_k$ $\neq$ $\mv{d}_k$ conditional on that $\mv{d}_k$ is transmitted by user $k$, and in (\ref{eq:SERs of 2 users}) we have assumed that all constellation symbols are selected for transmission with equal probability. In the rest of this paper, we consider the above SER upper bound as our performance metric.

To derive the PEP, we need to make one further assumption that the interference-plus-noise term $\mv{w}_k$ in (\ref{eq:output r_k 2}) is Gaussian distributed. This means that for deriving the PEP of user $k$, we need to assume that $\mv{d}_{l}$ $\sim\mathcal{N}(0,\mv{I})$, $\forall$ $l\in\{1, \dots, K\}$, $l\neq k$, i.e., $\mv{d}_l$ is a Gaussian random vector with zero mean and identity covariance matrix, despite of the practical $M_l$-ary modulation used. Under the above Gaussian assumption for the interference and hence the interference-plus-noise, after the application of the whitening filter, the interference-plus-noise $\mv{R}_k^T\mv{w}_k$ in (\ref{eq:output r_k 2}) becomes a Gaussian random vector with zero mean and identity covariance matrix. Then the optimal maximum likelihood (ML) detector for user $k$ can be shown to be equivalent to the Euclidean-distance based detector: for each symbol transmitted by user $k$, its receiver finds the constellation symbol $\mv{\hat{d}}_k$ which gives the smallest distance of $||\mv{r}_k-|h_{kk}|\mv{R}_k^T\mv{A}_k\mv{\hat{d}}_k||^2$ and declares it as the transmitted symbol. Thereby, we are able to obtain a closed-form expression for the PEP as shown in the following lemma.  

\begin{lemma}\label{lemma:prob d1-->tilde d1}
Assuming the interference-plus-noise vector $\mv{w}_k$ is Gaussian distributed and the Euclidean-distance based detector is used, the PEP at the receiver of user $k$ is given by
\begin{align}\label{eq:PEP optimal decoder}
Pr\{\mv{d}_k \to \mv{\tilde{d}}_k|\mv{d}_k\}  = Q\left(  \frac{ |h_{kk}| \sqrt{ (\mv{d}_k - \mv{\tilde{d}}_k)^T \mv{A}_k^T \mv{W}_k^{-1} \mv{A}_k (\mv{d}_k - \mv{\tilde{d}}_k) } }   {  2  }  \right),
\end{align}
where $Q(x)$ $\triangleq$ $\frac{1}{\sqrt{2\pi}}\int_{x}^{\infty}\exp(-u^2/2)du$.
\end{lemma} 
\begin{IEEEproof}
Please refer to Appendix \ref{proof:prob d1-->tilde d1}.
\end{IEEEproof}

\begin{figure*}
\begin{subfigure}[t]{1\textwidth}
 		\centering
 		\includegraphics[width=12.5cm, height=9.5cm]{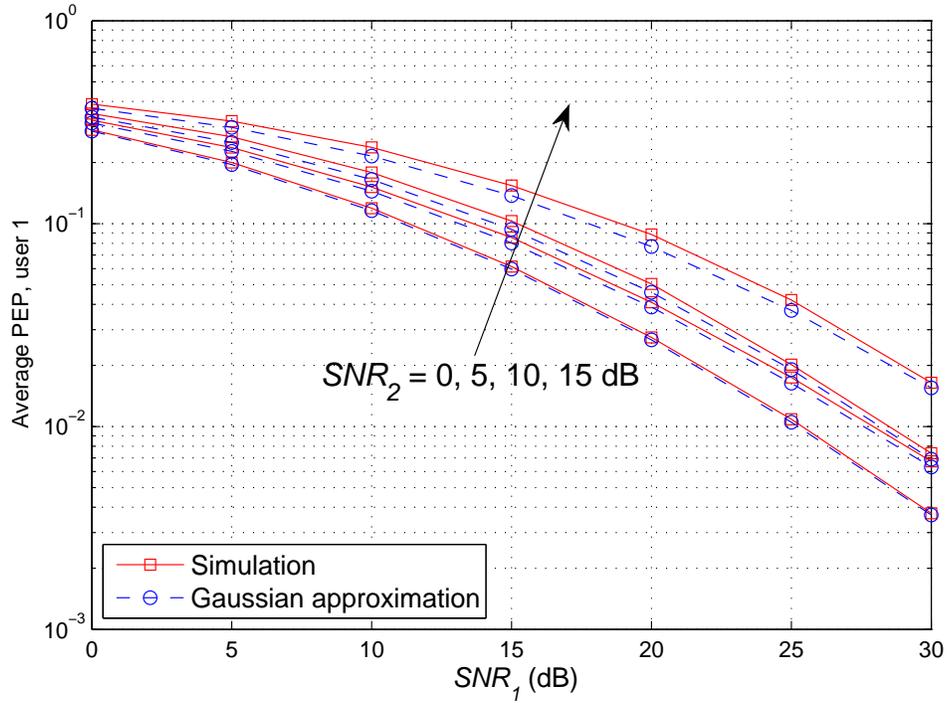}
 		\caption{}\label{fig:GA_SER_1}
\end{subfigure} \vspace{0.2in} \\
\begin{subfigure}[t]{1\textwidth}
 		\centering
 		\includegraphics[width=12.5cm,height=9.5cm]{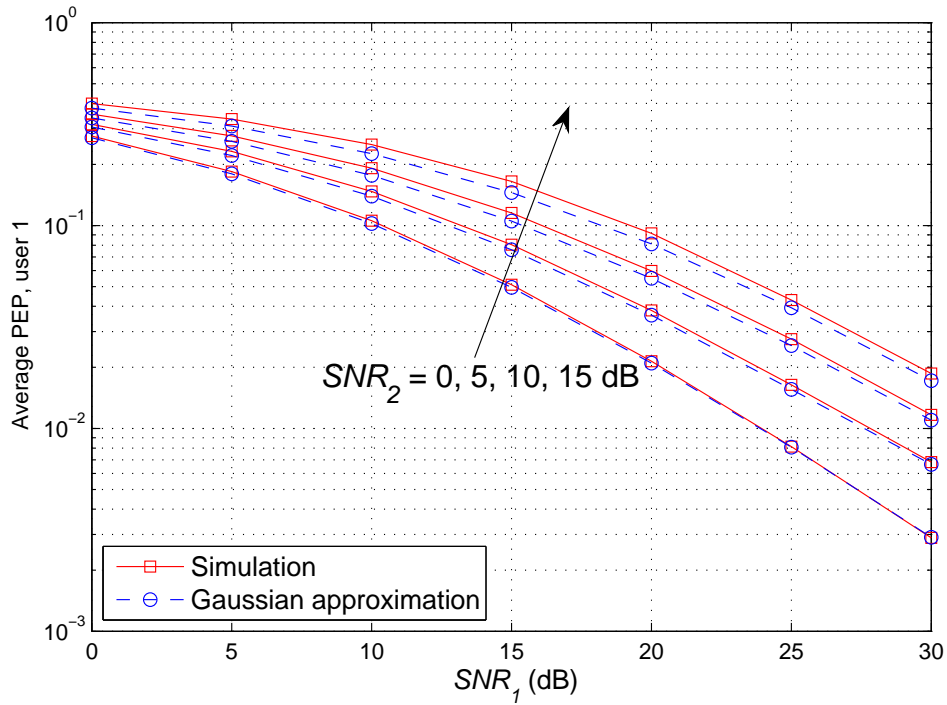}
 		\caption{}\label{fig:GA_SER_2}
\end{subfigure}
\captionsetup{width=0.65\textwidth}
\caption{Comparison between the simulation and analytical Gaussian approximated PEPs for user $1$ with 8PSK. The modulation of user 2 is either (a) QPSK or (b) 8PSK.}\label{fig:2}
\end{figure*}

In Fig. \ref{fig:2}, we verify the analytical result for the PEP in (\ref{eq:PEP optimal decoder}) for a two-user SISO IC by comparing the exact average PEP of user $1$ obtained by simulations versus $\text{SNR}_1$. Here we assume that user 1 applies 8PSK modulation, while user 2 employs either QPSK (Fig. \ref{fig:GA_SER_1}) or 8PSK (Fig. \ref{fig:GA_SER_2}). The channel is assumed to be Rayleigh fading, i.e., the channel coefficients $h_{kl}$'s are generated as i.i.d. $\mathcal{CN}(0,1)$, $k,l=1,2$. For each channel realization, the elements of $\mv{A}_k$ are randomly generated as real Gaussian RVs each distributed as $\mathcal{N}(0,1)$, $k=1,2$. $\mv{A}_k$ is then scaled so that $\mv{Tr}(\mv{A}_k \mv{A}_k^T)$ $= P_{k}$. Also, the SNR of user 2 is set as $\text{SNR}_2$ $\in$ $\{$0, 5, 10, 15$\}$ dB. The numbers of transmitted symbols and channel realizations are $10^6$ and 300, respectively. From Fig. \ref{fig:2}, we observe that the simulation results and the analytical results (\ref{eq:PEP optimal decoder}) based on Gaussian approximation are closely matched for all cases. Note that the PEPs are quite large here since the precoding matrices $\mv{A}_k$'s are randomly generated and are not optimized yet.

\section{Maximum PEP Minimization}\label{sec:min max PEP}

In order to minimize the maximum SER given in (\ref{eq:SERs of 2 users}) among all $K$ users with improper signalling, in this section we first propose an indirect approach by optimizing the precoding matrices $\mv{A}_k$'s of all users to minimize the maximum PEP given in (\ref{eq:PEP optimal decoder}) among all different transmitted symbols for each user $k$, as well as over all user $k$'s. Given a set of transmit power constraints for the users, $P_k$'s, the optimization problem is thus formulated as
\begin{align}
\mathrm{(P1a)}\mathop{\mathtt{min.}}_{ s, \big\{ \mv{A}_k \big\}_{k=1}^K} 
& ~~ s
\nonumber \\
\mathtt{s.t.} & ~~ Pr\{\mv{d}_k \to \mv{\tilde{d}}_k|\mv{d}_k\} \leq s, \notag \\
& ~~ \mv{Tr}(\mv{A}_k \mv{A}_k^T) \leq P_{k}, \notag \\
& ~~ \forall ~ \mv{\tilde{d}}_k \neq \mv{d}_k, ~ k=1, \dots, K. \notag
\end{align}

Using (\ref{eq:PEP optimal decoder}), problem (P1a) is equivalent to the following problem
\begin{align}\label{eq:P1b}
\mathrm{(P1b)}\mathop{\mathtt{max.}}_{ t, \big\{ \mv{A}_k \big\}_{k=1}^K} 
& ~t
\nonumber \\
\mathtt{s.t.} & ~ |h_{kk}|^2 (\mv{d}_k - \mv{\tilde{d}}_k)^T \mv{A}_k^T \mv{W}_k^{-1} \mv{A}_k (\mv{d}_k - \mv{\tilde{d}}_k) \geq t, \notag \\
& ~ \mv{Tr}(\mv{A}_k \mv{A}_k^T) \leq P_{k}, \notag \\
& ~ \forall ~ \mv{\tilde{d}}_k \neq \mv{d}_k, ~ k=1, \dots, K. \notag
\end{align}

Note that in the above problem, the first constraint for each user $k$ in fact corresponds to $D_k = \frac{M_k (M_k-1)}{2}$ number of constraints, which are for all the different signal constellation pairs, $\mv{\tilde{d}}_k$ $\neq$ $\mv{d}_k$. We denote $\mv{F}_k$ as a $D_k \times 2$ matrix which consists of all the vectors $\mv{d}_k - \mv{\tilde{d}}_k$, $\forall$ $\mv{\tilde{d}}_k$ $\neq$ $\mv{d}_k$. For example, the corresponding $\mv{F}_k$'s for the normalized QPSK and 8PSK signal constellations are
\begin{align}
\mv{F}_{k,QPSK} = \begin{bmatrix} 0 & 2 & 2 & 2 & 2 & 0 \\ 2 & 2 & 0 & 0 & -2 & -2 \end{bmatrix}^T,
\end{align}


\begin{align}
\mv{F}_{k,8PSK} = \left[ \begin{matrix} \sqrt{2}-1 & \sqrt{2} & \sqrt{2}+1 & 2\sqrt{2} & \sqrt{2}+1 & \sqrt{2} & \sqrt{2}-1 & 1 & 2 & \sqrt{2}+1 & 2 & 1 \notag \\ -1 & -\sqrt{2} & -1 & 0 & 1 & \sqrt{2} & 1 & -\sqrt{2}+1 & 0 & 1 & 2 & \sqrt{2}+1 \end{matrix} \right. \notag \\
\begin{matrix} 0 & 1 & \sqrt{2} & 1 & 0 & -1 & \sqrt{2}-1 & 0 & -1  \\
2 & \sqrt{2}-1 & \sqrt{2} & \sqrt{2}+1 & 2\sqrt{2} & \sqrt{2}+1 & 1 & 2 & \sqrt{2}+1 \end{matrix} \notag \\
\left. \begin{matrix}
-2 & -\sqrt{2}+1 & -\sqrt{2} & -\sqrt{2}-1 & -1 & -2 & -1 \\
2& 1 & \sqrt{2} & 1 & \sqrt{2}-1 & 0 & -\sqrt{2}+1
\end{matrix} \right]^T.
\end{align}
  
Then the first set of constraints in problem (P1b) for each user $k$ is equivalent to
\begin{align}\label{eq:user 1 constraints 2}
|h_{kk}|^2\left[ \mv{F}_k \mv{A}^T \mv{W}_k^{-1} \mv{A} \mv{F}_k^T \right]_{ii} \geq t, ~~ i = 1, \dots, D_k.
\end{align}

We have the following observations: 
\begin{enumerate}
\item Some rows of $\mv{F}_{k}$ are linearly dependent in a pair-wise manner (i.e., they are identical subject to a scaling multiplication).
\item Among any two pair-wise linearly dependent rows of $\mv{F}_{k}$, only the one with the smallest norm may correspond to an active constraint in (\ref{eq:user 1 constraints 2}).    
\end{enumerate}

Therefore, we can reduce the number of constraints in (\ref{eq:user 1 constraints 2}) by eliminating some rows in each $\mv{F}_k$ as follows. Denote $\mv{Q}_k$ as a $\bar{D}_k \times 2$ matrix consisting of the vectors $\mv{d}_k - \mv{\tilde{d}}_k$ with $\mv{\tilde{d}}_k$ $\neq$ $\mv{d}_k$, in which all the rows are pair-wise linearly independent. Furthermore, each row vector in $\mv{Q}_k$ has the smallest norm among all of its pair-wise linearly dependent vectors in $\mv{F}_k$. For example, the resulting matrices of $\mv{Q}_k$'s for normalized QPSK and 8PSK are given by
\begin{align}
\mv{Q}_{k,QPSK} = \begin{bmatrix} 0 & 2 & 2 & 2 \\ 2 & 2 & 0 & -2 \end{bmatrix}^T,
\end{align}

\begin{align}
\mv{Q}_{k,8PSK} = \begin{bmatrix} \sqrt{2}-1 & \sqrt{2} & 1 & 2 & 1 & \sqrt{2} & \sqrt{2}-1 & 0 \\ -1 & -\sqrt{2} & -\sqrt{2}+1 & 0 & \sqrt{2}-1 & \sqrt{2} & 1 & 2 \end{bmatrix}^T.
\end{align}

Compared to $\mv{F}_k$, the reduced matrix $\mv{Q}_k$ has significantly smaller number of rows. For example, $\bar{D}_{k,QPSK}$ = 4 and $\bar{D}_{k,8PSK}$ = 8 as compared to $D_{k,QPSK}$ = 6 and $D_{k,8PSK}$ = 28, respectively. Although applying $\mv{Q}_k$ instead of $\mv{F}_k$ to (\ref{eq:user 1 constraints 2}) does not essentially change the effectiveness of the constraints, it helps reduce the number of constraints for each user $k$ and hence the complexity for solving problem (P1b). This complexity saving is more significant with higher-order modulations. Problem (P1b) is accordingly reduced to
\begin{align}
\mathrm{(P1c)}:~\mathop{\mathtt{max.}}_{ t, \big\{ \mv{A}_k \big\}_{k=1}^K} 
& ~~ t
\nonumber \\
\mathtt{s.t.} & ~~ |h_{kk}|^2\left[ \mv{Q}_k \mv{A}_k^T \mv{W}_k^{-1} \mv{A}_k \mv{Q}_k^T \right]_{ii} \geq t, \notag \\
& ~~ \mv{Tr}(\mv{A}_k \mv{A}_k^T) \leq P_{k,T}, \notag \\
& ~~ i=1,\dots, \bar{D}_k, k=1, \dots, K. \notag
\end{align}

Problem (P1c) can be shown to be non-convex, and thus it is in general difficult to find the optimal solution for this problem. In the following, we propose an efficient algorithm that is guaranteed to find at least a locally optimal solution for problem (P1c). First, we introduce a set of auxiliary variables $\big\{\mv{B}_k\big\}_{k=1}^K$ and reformulate problem (P1c) as the following optimization problem.
\begin{proposition}\label{prop:P2a P2b}
Problem (P1c) is equivalent to the following problem
\begin{align}
\mathrm{(P1d)}:~\mathop{\mathtt{min.}}_{ \alpha, \big\{ \mv{A}_k \big\}_{k=1}^K, \big\{ \mv{B}_k \big\}_{k=1}^K} 
& ~~ \alpha
\nonumber \\
\mathtt{s.t.} & ~~ \left[\mv{G}_k \right]_{ii} \leq \alpha, \notag \\
& ~~ \mv{Tr}(\mv{A}_k \mv{A}_k^T) \leq P_{k}, \notag \\
& ~~ i=1,\dots, \bar{D}_k, ~ k=1, \dots, K. \notag
\end{align}
in which
\begin{align}\label{eq:E1E2}
\mv{G}_k = \sum_{l=1, l\neq k}^K |h_{kl}|^2 \mv{B}_k^T \mv{J}(\phi_{kl}) \mv{A}_l \mv{A}_l^T \mv{J}^T(\phi_{kl}) \mv{B}_l - |h_{kk}| \mv{B}_k^T \mv{A}_k \mv{Q}_k^T - |h_{kk}| \mv{Q}_k \mv{A}_k^T \mv{B}_k + \frac{\sigma_k^2}{2} \mv{B}_k^T \mv{B}_k.
\end{align}
\end{proposition}
\begin{IEEEproof}
Please refer to Appendix \ref{proof:P2a P2b}.
\end{IEEEproof}

Note that
\begin{align}\label{eq:E1E2_ii}
& \left[ \mv{G}_k \right]_{ii} =  \frac{\sigma_k^2}{2} ||\mv{b}_{k,i}||^2 + \sum_{l=1, l\neq k}^K |h_{kl}|^2 ||\mv{b}_{k,i}^T \mv{J}(\phi_{kl}) \mv{a}_{l,1}||^2 \notag \\
& \qquad \qquad \qquad \qquad + \sum_{l=1, l\neq k}^K |h_{kl}|^2 ||\mv{b}_{k,i}^T \mv{J}(\phi_{kl}) \mv{a}_{l,2}||^2 - 2|h_{kk}|\mv{Tr}\left(\mv{q}_{k,i}^T \mv{b}_{k,i}^T \mv{A}_k  \right),
\end{align}
where $\mv{b}_{k,i}$ is the $i$-th column of the matrix $\mv{B}_k$; $\mv{a}_{k,1}$ and $\mv{a}_{k,2}$ are the first and second columns of matrix $\mv{A}_k$; and $\mv{q}_{k,i}$ is the $i$-th row of matrix $\mv{Q}_k$, $k=1, \dots, K$. 

\begin{table}[t]
\centering
\caption{\textsc{Minmax-PEP Algorithm}}\label{tab:Minmax-PEP}
\begin{tabular}{l l}
\hline
\hline
\vspace{-0.0in} & \\
1. & Initialize $\mv{A}_k$ $=$ $\sqrt{P_k/2} \mv{I}$, $k=1,\dots,K$. \\
2. & Set $\mv{B}_k^T$ = $|h_{kk}|$ $\mv{Q}_k$ $\mv{A}_k^T$ $\Big( \frac{\sigma_k^2}{2}\mv{I} + \sum_{l=1, l\neq k}^K |h_{kl}|^2$ $\mv{J}(\phi_{kl}) \mv{A}_l \mv{A}_l^T \mv{J}^T(\phi_{kl}) \Big)^{-1}$, $k=1, \dots, K$. \\
3. & Update $\big\{ \mv{A}_k \big\}_{k=1}^K$ by solving problem (Ps1). \\
4. & Repeat steps 2 and 3 until both $\big\{ \mv{A}_k \big\}_{k=1}^K$ and $\big\{ \mv{B}_k \big\}_{k=1}^K$ converge within the prescribed accuracy. \vspace{+0.1in}\\
\hline
\end{tabular} \vspace{-0.0in}
\end{table}

Since problem (P1d) can be shown to be  convex over each of the two sets $\big\{ \mv{A}_k \big\}_{k=1}^K$ and $\big\{ \mv{B}_k \big\}_{k=1}^K$ when one of them is given as fixed, we can apply the technique of alternating optimization to solve the problem iteratively. Although global convergence is not guaranteed in general, this approach ensures local convergence and often leads to a good suboptimal solution when initialized properly. Specifically, given $\big\{ \mv{A}_k \big\}_{k=1}^K$, the solution of $\big\{ \mv{B}_k \big\}_{k=1}^K$ can be already deduced from the proof of Proposition \ref{prop:P2a P2b} (see Appendix \ref{proof:P2a P2b}). On the other hand, optimizing $\big\{ \mv{A}_k \big\}_{k=1}^K$ with given $\big\{ \mv{B}_k \big\}_{k=1}^K$ can be obtained by solving the following convex problem (Ps1) by applying the primal-dual interior point method \cite{Boyd01}, via existing softwares, e.g., CVX \cite{Grant01}.
\begin{align}
\mathrm{(Ps1)}:~\mathop{\mathtt{min.}}_{ \alpha, \big\{ \mv{A}_k \big\}_{k=1}^K } 
& ~~ \alpha
\nonumber \\
\mathtt{s.t.}& ~~ \frac{\sigma_k^2}{2} ||\mv{b}_{k,i}||^2 \sum_{l=1, l\neq k}^K |h_{kl}|^2 ||\mv{b}_{k,i}^T \mv{J}(\phi_{kl}) \mv{a}_{l,1}||^2 \notag \\
& \qquad \qquad \qquad \qquad + \sum_{l=1, l\neq k}^K |h_{kl}|^2 ||\mv{b}_{k,i}^T \mv{J}(\phi_{kl}) \mv{a}_{l,2}||^2 - 2 |h_{kk}| \mv{Tr}\left(\mv{q}_{k,i}^T \mv{b}_{k,i}^T \mv{A}_k  \right) \leq \alpha, \notag \\
& ~~ \mv{Tr}(\mv{A}_k \mv{A}_k^T) \leq P_{k}, \notag \\
& ~~ k=1, \dots, K, ~~ i=1,\dots, \bar{D}_k. \notag
\end{align}

To summarize, the proposed algorithm to solve problem (P1d) and hence (P1a) is given in Table \ref{tab:Minmax-PEP}, which is referred to as Minmax-PEP. Note that in Table I, we have initialized $\mv{A}_k$ $=$ $\sqrt{P_k/2} \mv{I}$, $k=1,\dots,K$, i.e., assuming all users to employ conventional proper signalling initially. We further note that the Minmax-PEP algorithm fully exploits the CSI and constellation information of the users, represented by the channel coefficients $h_{kl}$'s and matrices $\big\{ \mv{Q}_k \big\}_{k=1}^K$, respectively.


\section{Maximum SER Minimization}\label{sec:min SER upper-bound}

The Minmax-PEP algorithm proposed in Section \ref{sec:min max PEP} minimizes the worst PEP for every symbol pair of the users. In this section, we directly minimize the SERs given in (\ref{eq:SERs of 2 users}), and refer to the resulting algorithm as Minmax-SER. It is expected that Minmax-SER, by further balancing the PEPs for each of the users, is able to achieve a better SER performance than Minmax-PEP at a cost of higher complexity for optimization.

Specifically, we aim to minimize the maximum of $\text{SER}_k$'s which are given in (\ref{eq:SERs of 2 users}), $k=1, \dots, K$. Using Lemma \ref{lemma:prob d1-->tilde d1}, the optimization problem is formulated as
\begin{align}
\mathrm{(P2a)}:~\mathop{\mathtt{min.}}_{ t, \big\{ \mv{A}_k \big\}_{k=1}^K} 
& ~~ t
\nonumber \\
\mathtt{s.t.} & ~~ \frac{1}{M_k(M_k-1)} \sum_{\mv{d}_k} \sum_{\mv{\tilde{d}}_k \neq \mv{d}_k} Q\left(  \frac{ |h_{kk}| \sqrt{ (\mv{d}_k - \mv{\tilde{d}}_k)^T \mv{A}_k^T \mv{W}_k^{-1} \mv{A}_k (\mv{d}_k - \mv{\tilde{d}}_k) } }   {  2  }  \right) \leq t, \notag \\
& ~~ \mv{Tr}(\mv{A}_k \mv{A}_k^T) \leq P_{k}, \notag \\
& ~~ \forall ~ \mv{\tilde{d}}_k \neq \mv{d}_k, k=1, \dots, K. \notag
\end{align}

We then state the following proposition.
\begin{proposition}\label{prop:P4a}
Problem (P2a) is equivalent to the following problem.
\begin{align}
\mathrm{(P2b)}:~\mathop{\mathtt{min.}}_{ t, \big\{ \mv{A}_k \big\}_{k=1}^K, \big\{ \mv{B}_k \big\}_{k=1}^K, \big\{ t_{k,i} \big\}_{ \substack{k=1, \dots, K \\ i=1, \dots, D_k} } } 
& ~~ t
\nonumber \\
\mathtt{s.t.} & ~~ \left[ \mv{S}_k \right]_{ii} \leq - t_{k,i}^2, \notag \\
& ~~ \frac{1}{D_k} \sum_{i=1}^{D_k} Q\left(  \frac{t_{k,i}}{2}  \right) \leq t, \notag \\
& ~~ t_{k,i} \geq 0, \notag \\
& ~~ \mv{Tr}(\mv{A}_k \mv{A}_k^T) \leq P_{k}, \notag \\
& ~~ k=1, \dots, K, ~~ i=1,\dots, D_k. \notag
\end{align}
in which
\begin{align}\label{eq:S1S2}
\mv{S}_k = \sum_{l=1, l\neq k}^K |h_{kl}|^2 \mv{B}_k^T \mv{J}(\phi_{kl}) \mv{A}_l  \mv{A}_l^T \mv{J}^T(\phi_{kl}) \mv{B}_l - |h_{kk}| \mv{B}_k^T \mv{A}_k \mv{F}_k^T - |h_{kk}| \mv{F}_k \mv{A}_k^T \mv{B}_k + \frac{\sigma_k^2}{2} \mv{B}_k^T \mv{B}_k.
\end{align}
\end{proposition}
\begin{IEEEproof}
The proof is similar to that of Proposition \ref{prop:P2a P2b} and is thus omitted for brevity.
\end{IEEEproof}

Similarly as for the matrix $\mv{G}_k$ defined in Section \ref{sec:min max PEP}, we have
\begin{align}
& \left[ \mv{S}_k \right]_{ii} =  \frac{\sigma_k^2}{2} ||\mv{b}_{k,i}||^2 + \sum_{l=1, l\neq k}^K |h_{kl}|^2 ||\mv{b}_{k,i}^T \mv{J}(\phi_{kl}) \mv{a}_{l,1}||^2 \notag \\
& \qquad \qquad \qquad \qquad + \sum_{l=1, l\neq k}^K |h_{kl}|^2 ||\mv{b}_{k,i}^T \mv{J}(\phi_{kl}) \mv{a}_{l,2}||^2 - 2|h_{kk}|\mv{Tr}\left(\mv{f}_{k,i}^T \mv{b}_{k,i}^T \mv{A}_k  \right),
\end{align}
where $\mv{f}_{k,i}$ is the $i$-th row of the matrix $\mv{F}_k$, $k=1, \dots, K$.  
 
We again employ alternating minimization to solve (P2b). Accordingly, we propose the Minmax-SER algorithm given in Table \ref{tab:Minmax-SER}, in which problem (Ps2) is defined as
\begin{align}
\mathrm{(Ps2)}:~\mathop{\mathtt{min.}}_{ t, \big\{ \mv{A}_k \big\}_{k=1}^K, \big\{ t_{k,i} \big\}_{ \substack{k=1, \dots, K \\ i=1, \dots, D_k} } } 
& ~~ \alpha
\nonumber \\
\mathtt{s.t.} & ~~ \frac{\sigma_k^2}{2} ||\mv{b}_{k,i}||^2 + \sum_{l=1, l\neq k}^K |h_{kl}|^2 ||\mv{b}_{k,i}^T \mv{J}(\phi_{kl}) \mv{a}_{l,1}||^2 \notag \\
& \quad + \sum_{l=1, l\neq k}^K |h_{kl}|^2 ||\mv{b}_{k,i}^T \mv{J}(\phi_{kl}) \mv{a}_{l,2}||^2 - 2 |h_{kk}| \mv{Tr}\left(\mv{f}_{k,i}^T \mv{b}_{k,i}^T \mv{A}_k  \right) \leq -t_{k,i}^2, \notag \\
& ~~ \frac{1}{D_k} \sum_{i=1}^{D_k} Q\left(  \frac{t_{k,i}}{2}  \right) \leq t, \notag \\
& ~~ t_{k,i} \geq 0, \notag \\
& ~~ \mv{Tr}(\mv{A}_k \mv{A}_k^T) \leq P_{k}, \notag \\
& ~~ k=1, \dots, K, ~~ i=1,\dots, D_k. \notag 
\end{align} 

\begin{table}[t]
\centering
\caption{\textsc{Minmax-SER Algorithm}}\label{tab:Minmax-SER}
\begin{tabular}{l l}
\hline
\hline
\vspace{-0.0in} & \\
1. & Initialize $\mv{A}_k$ $=$ $\sqrt{P_k/2} \mv{I}$, $k=1,\dots,K$. \\
2. & Set $\mv{B}_k^T$ = $|h_{kk}|$ $\mv{F}_k$ $\mv{A}_k^T$ $\Big( \frac{\sigma_k^2}{2}\mv{I} + \sum_{l=1, l\neq k}^K |h_{kl}|^2$ $\mv{J}(\phi_{kl}) \mv{A}_l \mv{A}_l^T \mv{J}^T(\phi_{kl}) \Big)^{-1}$, $k=1, \dots, K$. \\
3. & Update $\big\{ \mv{A}_k \big\}_{k=1}^K$ by solving problem (Ps2). \\
4. & Repeat steps 2 and 3 until both $\big\{ \mv{A}_k \big\}_{k=1}^K$ and $\big\{ \mv{B}_k \big\}_{k=1}^K$ converge within the prescribed accuracy. \vspace{+0.1in}\\
\hline
\end{tabular} \vspace{-0.0in}
\end{table}

Note that the Q-function $Q(x)$ is convex and strictly decreasing in the domain $x\in[0, +\infty)$. Therefore (Ps2) is a convex optimization problem and can be solved efficiently by, e.g., the primal-dual interior-point method \cite{Boyd01}. Similar to the Minmax-PEP, we observe that the Minmax-SER fully exploits the CSI and constellation information of the users, represented by the channel coefficients $h_{kl}$'s and matrices $\big\{ \mv{F}_k \big\}_{k=1}^K$, respectively. It is worth noting that the Minmax-SER algorithm here requires the knowledge of every row of $\mv{F}_k$'s, and therefore the reduced matrices $\mv{Q}_k$'s for Minmax-PEP algorithm in Section \ref{sec:min max PEP} cannot be used in this case. However, similar to Minmax-PEP algorithm, Minmax-SER is also guaranteed to converge to (at least) a local optimal solution.
  
\section{Benchmark Schemes}\label{sec:benchmark}

In this section, we present alternative approaches in the existing literature to optimize the precoding and decoding matrices, i.e., $\big\{ \mv{A}_k \big\}_{k=1}^K$ and $\big\{ \mv{R}_k \big\}_{k=1}^K$ in (\ref{eq:output r_k}), for the $K$-user SISO-IC with proper/improper signalling. It is worth pointing out that, unlike our proposed Minmax-PEP/SER algorithms, these benchmark schemes might not be originally designed for minimizing the SER in ICs directly. 

\subsection{Proper Signaling with Power Control}
For the $K$-user SISO IC with conventional proper signaling, the precoding and decoding matrices are reduced to $\mv{A}_k = \sqrt{\frac{p_{k}}{2}} \mv{I}$ and $\mv{R}_k = \mv{I}$, where $p_k \leq P_{k}$. Considering the use of ML or Euclidean-distance based detection at each receiver, user $k$ detects that the symbol $\mv{\hat{d}}_k$ is transmitted if it gives the smallest distance of $||\mv{r}_k-\frac{p_{k}|h_{kk}|}{2}\mv{\hat{d}}_k||^2$ among all signal symbols. In this case, we search over all user power allocations $(p_1, \dots, p_K)$ subject to $p_k\leq P_{k}$ to find the best $(p_1, \dots, p_K)$ which achieves the minimum of $\max \text{SER}_k$, $k=1,\dots,K$. We refer to this scheme as Proper Signalling with Power Control (PS-PC).

\subsection{MSE-based Schemes}\label{sec:min max MSE per stream}

Although there has been no existing study on directly minimizing the SERs for ICs with improper signalling, minimizing the MSEs has been considered as an alternative approach (see, e.g., \cite{Shen01}, \cite{Chen01}). Following \cite{Shen01}, \cite{Chen01}, we first define the MSE matrix for $\mv{r}_k$ as
\begin{align}
\mv{E}_k & = \mathbb{E}\left[ (\mv{r}_k-\mv{d}_k) (\mv{r}_k-\mv{d}_k)^T \right] \notag \\
& = \mv{R}_k^T \Big( \frac{\sigma_k^2}{2}\mv{I} + \sum_{l=1, l\neq k}^K |h_{kl}|^2 \mv{J}(\phi_{kl}) \mv{A}_l  \mv{A}_l^T \mv{J}^T(\phi_{kl}) + |h_{kk}|^2 \mv{A}_k  \mv{A}_k^T \Big) \mv{R}_k - |h_{kk}| \mv{R}_k^T\mv{A}_k - |h_{kk}| \mv{A}_k^T\mv{R}_k + \mv{I}, \notag
\end{align}
where we have used the following assumptions: $\mathbb{E}[\mv{d}_l\mv{d}_k^T] = \mv{0}$; $\mathbb{E}[\mv{d}_l\mv{n}_k^T] = \mv{0}$; and $\mathbb{E}[\mv{d}_k\mv{d}_k^T] = \mv{I}$. The minimization of the sum of MSEs and the maximum per-stream MSE can then be expressed as the following optimization problems in (\ref{eq:P3a}) and (\ref{eq:P3b}), respectively \cite{Shen01}, \cite{Chen01}.
\begin{align}\label{eq:P3a}
\mathop{\mathtt{min.}}_{ \big\{ \mv{A}_k \big\}_{k=1}^K, \big\{ \mv{R}_k \big\}_{k=1}^K } 
& ~~ \sum_{k=1}^K \mv{Tr}\left\{ \mv{E}_k \right\}
\nonumber \\
\mathtt{s.t.} & ~ \mv{Tr}(\mv{A}_k  \mv{A}_k^T) \leq P_{k}, ~ k=1, \dots, K.
\end{align}
\begin{align}\label{eq:P3b}
\mathop{\mathtt{min.}}_{ \big\{ \mv{A}_k \big\}_{k=1}^K, \big\{ \mv{R}_k \big\}_{k=1}^K } 
& ~~ \max_{k=1, \dots, K} \max_{i=1, 2} ~~ \left[ \mv{E}_k \right]_{i,i}
\nonumber \\
\mathtt{s.t.} & ~ \mv{Tr}(\mv{A}_k  \mv{A}_k^T) \leq P_{k}, ~ k=1, \dots, K.
\end{align}

The algorithms for solving the above two problems have been given in \cite{Shen01}, \cite{Chen01} and are thus omitted here for brevity. We denote the algorithms to solve (\ref{eq:P3a}) and (\ref{eq:P3b}) as Minsum-MSE and Minmax-MSE, respectively. Note that the convergence of these algorithms to (at least) a local optimum of (\ref{eq:P3a}) or (\ref{eq:P3b}) is guaranteed. Applying the obtained precoding and decoding matrices, the receiver of each user $k$ finds the nearest constellation symbol $\mv{\hat{d}}_k$ to $\mv{r}_k$ given in (\ref{eq:output r_k 2}), i.e., $\mv{\hat{d}}_k$ $=\arg \min_{\mv{d}_k}$ $||\mv{r}_k-\mv{d}_k||^2$, and then declares it as the transmitted symbol. 

\begin{remark}
It is worth noting that since the Minsum-MSE and Minmax-MSE only consider the MSE criterion, they therefore have not made use of the constellation information of the users. That is, the schemes exploit only the CSI but not the constellation information of the users. A direct consequence is that the precoding and decoding matrices designed for a particular channel realization are identical irrespective of users' modulations. It is thus expected that the Minsum-MSE and Minmax-MSE schemes might not perform as well as our proposed Minmax-PEP/SER algorithms, especially when the users employ higher-order modulations which are more susceptible to interference.
\end{remark}

\subsection{IA-based Schemes}\label{sec:IL IA}

It is known that the maximum sum-rate DoF of the equivalent $K$-user $2\times 2$ real IC given in (\ref{eq:real-valued IC}) with improper signalling is 2 (or equivalently 1 per complex dimension) when $K=2$, which is achieved when each of the two users sends one data stream, based on the principle of zero-forcing (ZF) \cite{Jafar01}. In this subsection, we consider another possible strategy to minimize the SER by applying IA-based precoder and decoder designs for the $K$-user SISO-IC with improper signalling. Specifically, we aim to find $\left\{ \mv{A}_k \right\}_{k=1}^K$ and $\left\{ \mv{R}_k \right\}_{k=1}^K$ such that \cite{Gomadam01}, \cite{Peters01}
\begin{align}\label{eq:IA conditions}
\mv{R}_k^T \mv{J}(\phi_{kl}) \mv{A}_{l} = 0, \notag \\
\text{rank}(\mv{R}_k^T \mv{A}_k) = 1.
\end{align}
in which $k,l$ $\in$ $\{1, \dots, K\}$, $l\neq k$. In (\ref{eq:IA conditions}), the first and second set of equations are for nullifying the interference and enforcing the desired signal to span exactly one dimension at each receiver, respectively. In this paper, we consider two well-established IA-based schemes, i.e., interference leakage minimization and SINR maximization \cite{Gomadam01}, \cite{Peters01}, denoted as MinIL-IA and MaxSINR-IA, respectively. 

Note that the above IA-based algorithms will yield the precoding and decoding matrices simplified as $2\times 1$ transmit and receive beamforming vectors, denoted by $\{\mv{v}_k\}_{k=1}^K$ and $\{\mv{u}_k\}_{k=1}^K$, respectively; thus, each user needs to transmit with only one-dimensional (1D) signal constellations. Therefore, for a fair comparison with other improper signalling schemes considered in this paper which use two-dimensional (2D) signal constellations, the transmitted symbols of IA-based schemes are assumed to be drawn from the 1D PAM modulation with the same number of symbols and average transmit power as the comparing 2D modulation schemes. For example, the corresponding constellations for the normalized QPSK and 8PSK are the normalized 4PAM and 8PAM, respectively, i.e.,
\begin{itemize}
\item QPSK: $\left\{ \begin{bmatrix} 1 \\ 1\end{bmatrix}, \begin{bmatrix} 1 \\ -1\end{bmatrix}, \begin{bmatrix} -1 \\ -1\end{bmatrix}, \begin{bmatrix} -1 \\ 1\end{bmatrix} \right\}$ $\xrightarrow{\quad}$ 4PAM: $\left\{ -3\sqrt{\frac{2}{5}}, -\sqrt{\frac{2}{5}}, \sqrt{\frac{2}{5}}, 3\sqrt{\frac{2}{5}} \right\}$.
\item 8PSK: $\left\{ \begin{bmatrix} 0 \\ \sqrt{2} \end{bmatrix}, \begin{bmatrix} 1 \\ 1\end{bmatrix}, \begin{bmatrix} \sqrt{2} \\ 0\end{bmatrix}, \begin{bmatrix} 1 \\ -1\end{bmatrix}, \begin{bmatrix} 0 \\ -\sqrt{2} \end{bmatrix}, \begin{bmatrix} -1 \\ -1\end{bmatrix}, \begin{bmatrix} -\sqrt{2} \\ 0\end{bmatrix}, \begin{bmatrix} -1 \\ 1\end{bmatrix} \right\}$ \\ \\ $\xrightarrow{\quad}$ 8PAM: $\left\{ -7\sqrt{\frac{2}{21}}, -5\sqrt{\frac{2}{21}}, -3\sqrt{\frac{2}{21}}, -\sqrt{\frac{2}{21}}, \sqrt{\frac{2}{21}}, 3\sqrt{\frac{2}{21}}, 5\sqrt{\frac{2}{21}}, 7\sqrt{\frac{2}{21}} \right\}$.
\end{itemize}

\begin{remark}
The design objective of the MinIL-IA and MaxSINR-IA is not the PEP or SER. As a consequence, the error rate performance of the IA-based schemes is also unpredictable and unguaranteed, similarly to the MSE-based schemes. Moreover, we note that the IA-based schemes do not take into account the constellation information of the users.
\end{remark}

\section{Numerical results}\label{sec:numerical results}

In this section, we compare the performance of our proposed Minmax-PEP/SER algorithms with the benchmark PS-PC, MSE-based, and IA-based schemes under both AWGN channel and cellular system setups. Note that for each channel realization, we first obtain the precoding and decoding matrices for each scheme and then apply them to find the average SER over $10^8$ number of randomly generated transmitted symbols by simulation.  

\subsection{AWGN Channel}\label{subsec:awgn}
\begin{figure}[t]
 		\centering 
 		\epsfxsize=0.85\linewidth
 		\captionsetup{width=0.85\textwidth} 
 		\includegraphics[width=14.5cm, height=10.5cm]{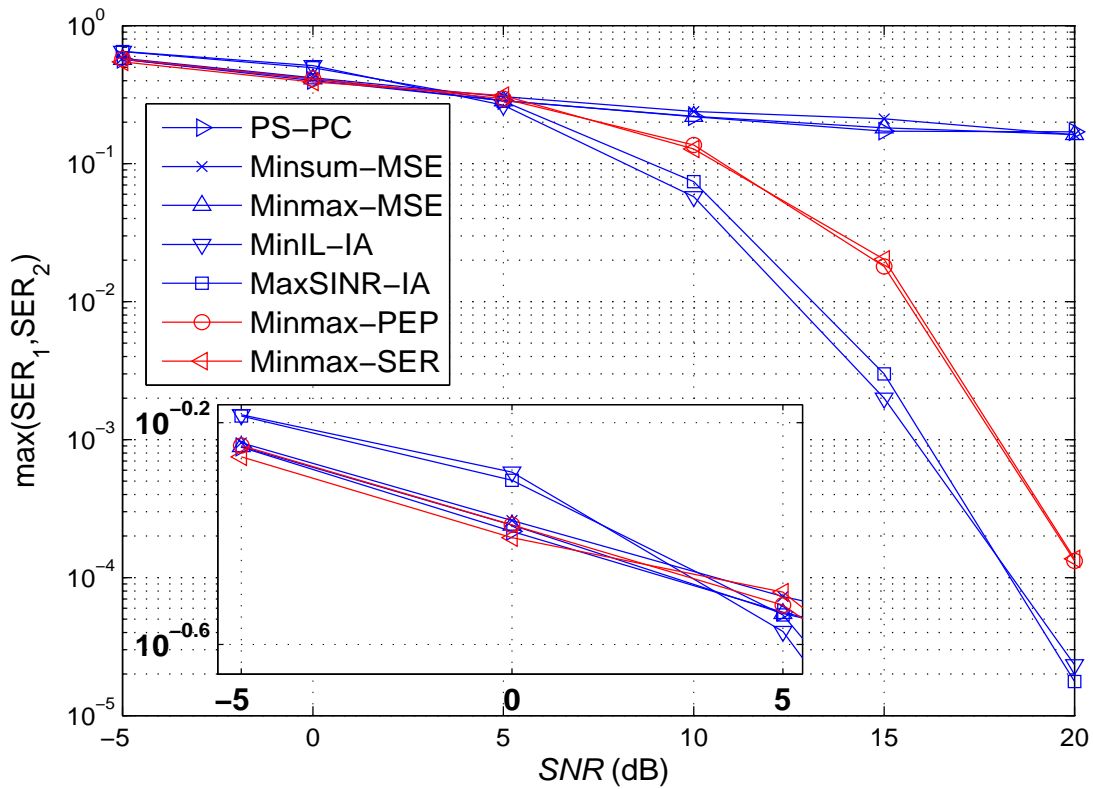}
 		\caption{Comparison of the minimized maximum SER over SNR in a two-user IC. \vspace{-0.0in}}\label{fig:2IC_SERfig1}
\end{figure}

In this subsection, we present simulation results to compare the SER performance of the proposed Minmax-PEP/SER with other benchmark schemes in $K$-user AWGN SISO-ICs. First, Fig. \ref{fig:2IC_SERfig1} shows the results for the minimized maximum SER of the users over SNR in a two-user IC with the channel coefficients given by   
\begin{align}
\begin{bmatrix} 1.9310e^{-j2.0228} & 0.7732e^{j0.5865} \\ 0.9249e^{j3.0213} & 2.3742e^{j0.2089}\end{bmatrix}. \notag
\end{align}

Here, we assume that both users apply 8PSK modulation. Therefore, the equivalent constellation for IA-based schemes is 8PAM. The SNRs of the two users are assumed to be equal and are varied from 0 to 20 dB. We observe that the PS-PC and MSE-based schemes result in saturated SERs for both users over SNR. This is because under such schemes, both users' signals span two dimensions each, and thus the SER performance is interference-limited with increasing SNR. 

In contrast, the maximum SERs by the proposed Minmax-PEP/SER and IA-based schemes are observed to decrease over SNR. Similar to the IA-based schemes, as the SNR increases, the precoding matrices $\mv{A}_1$ and $\mv{A}_2$ obtained from Minmax-PEP/SER gradually converge to rank-1 matrices, i.e., each user's precoded signal spans over only one dimension although the original signal $\mv{d}_k$ is drawn from a 2D constellation (8PSK). As a result, orthogonal transmissions of the two users are achieved at each receiver and the system SER is not limited by the interference as SNR increases. It is also observed that Minmax-PEP/SER perform better than IA-based schemes when SNR is less than $0$ dB; however, as SNR further increases, the minimized maximum SER of Minmax-PEP/SER becomes worse than that of IA-based schemes, which is mainly due to the use of suboptimal noise whitening filters for Minmax-PEP/SER in order to derive the closed-form PEP expression given in (\ref{eq:PEP optimal decoder}). Thus, it is interesting to jointly optimize the decoding and precoding matrices to further improve the SER performance in our proposed improper signaling schemes, which will be left to our future work. 

\begin{figure}[t]
 		\centering 
 		\epsfxsize=0.85\linewidth
 		\captionsetup{width=0.85\textwidth} 
 		\includegraphics[width=14.5cm, height=10.5cm]{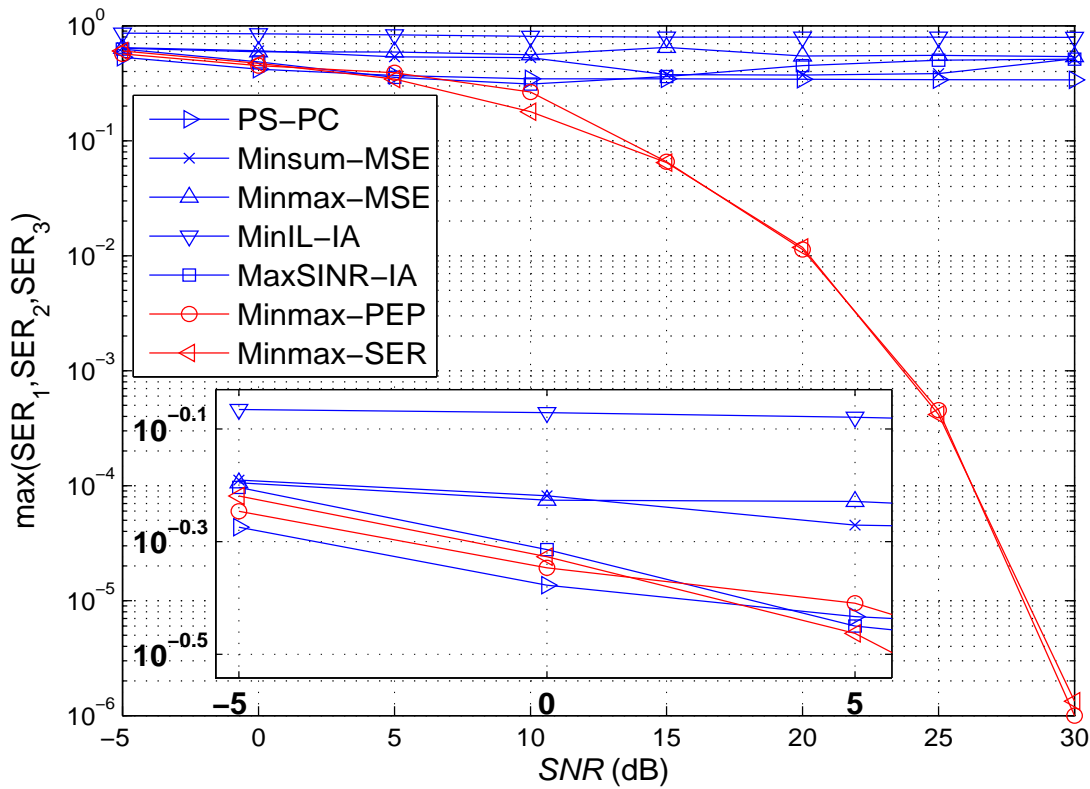}
 		\caption{Comparison of the minimized maximum SER over SNR in a three-user IC. \vspace{-0.0in}}\label{fig:3IC_SERfig1}
\end{figure}

Next, Fig. \ref{fig:3IC_SERfig1} compares the minimized maximum SER in a three-user IC given by
\begin{align}
\begin{bmatrix} 1.9310e^{-j2.0228} & 0.7732e^{j0.5865} & 0.9766e^{j1.1907} \\ 0.9249e^{j3.0213} & 2.3742e^{j0.2089} & 0.3009e^{-j1.5307} \\ 1.7628e^{-j0.4282} & 0.3127e^{-j1.4959} & 2.1935e^{j1.7364}\end{bmatrix}. \notag
\end{align}

Here users $(1,2,3)$ apply (QPSK, 8PSK, 8PSK) modulations, respectively. The equivalent constellations for IA-based schemes are hence (4PAM, 8PAM, 8PAM). The users' SNRs are set to be equal and are varied from 0 to 30 dB. It is observed that Minmax-PEP/SER achieves substantial SER improvement over all other schemes. Note that in this case, the IA-based schemes cannot reduce SER over SNR, since there is no feasible IA solution to (\ref{eq:IA conditions}) with given 1D signal constellations for all the three users. In Fig. \ref{fig:3IC_SERfig1}, we notice that Minmax-SER achieves a (slightly) better minimized SER than Minmax-PEP. However, the improvement of Minmax-SER over Minmax-PEP is observed only in the medium-SNR regime (0 to 15 dB). It is expected, since the solutions of both schemes in the low and high-SNR regimes are proper signaling and rank-1 precoding matrices, respectively. Thus the SER results of Minmax-SER and Minmax-PEP are identical under low or high-SNR regime.      

\subsection{Cellular System}\label{subsec:cellular}

\begin{figure*}

\begin{subfigure}[t]{0.5\textwidth}
 		\centering
 		\includegraphics[width=6.0cm,height=4.0cm]{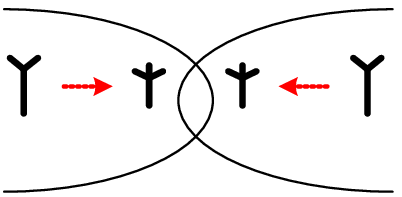}
 		\caption{Two cells.}\label{fig:2IC fading}
\end{subfigure}
\begin{subfigure}[t]{0.5\textwidth}
 		\centering
 		\includegraphics[width=6.0cm,height=5.0cm]{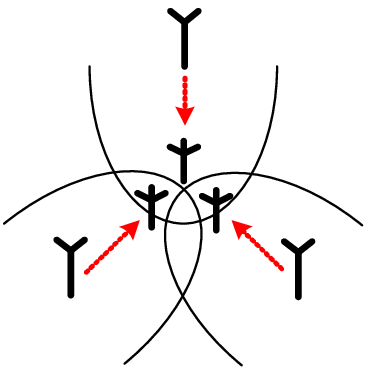}
 		\caption{Three cells.}\label{fig:3IC fading}
\end{subfigure}
\caption{Cellular system setups with parameters given in Table \ref{tab:cellular system}. \vspace{+0.2in} }\label{fig:fading}
\end{figure*}

\begin{figure}[t]
 		\centering 
 		\epsfxsize=0.48\linewidth
 		\captionsetup{width=0.75\textwidth} 
 		\includegraphics[width=14.5cm, height=10.5cm]{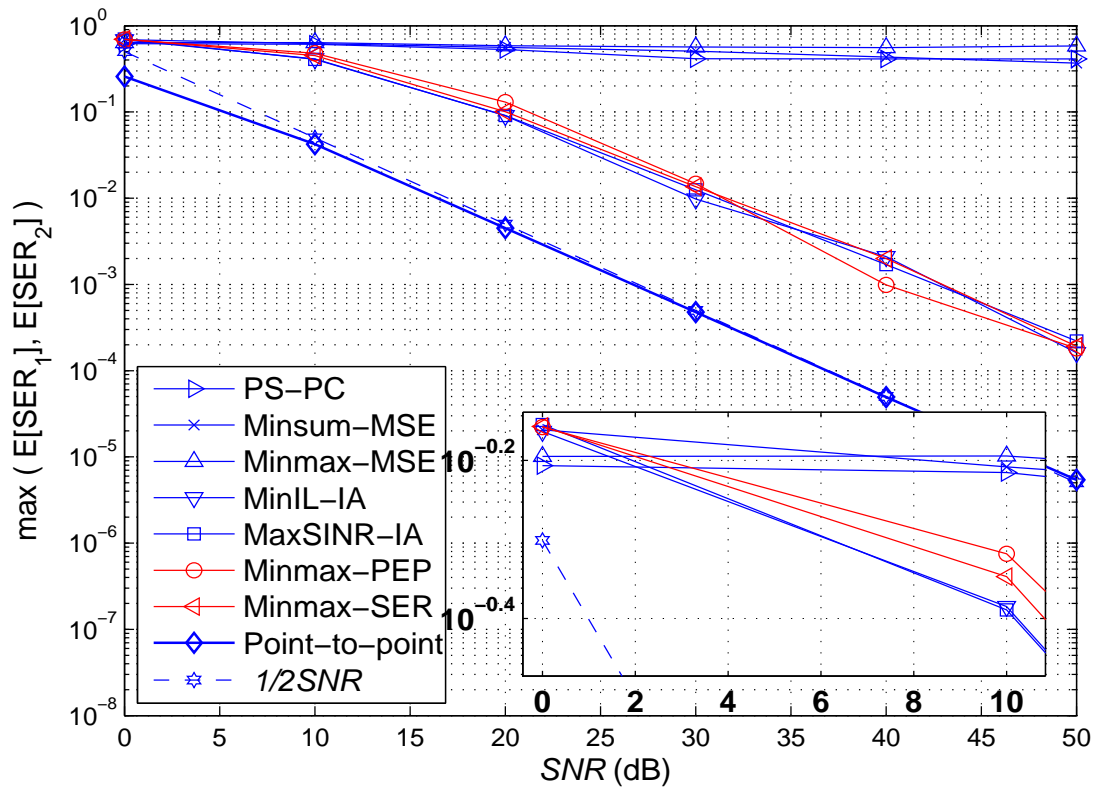}
 		\caption{The SER performance comparison in a cellular network with 2 cell-edge users. }\label{fig:2IC_SER fading}
\end{figure}

In this subsection, we compare the worst average SER performance of the Minmax-PEP/SER and other schemes under practical cellular system setups. We first assume two cells with two corresponding cell-edge users as shown in Fig. \ref{fig:2IC fading}, and compare the maximum average SERs of various schemes in Fig. \ref{fig:2IC_SER fading}. The system parameters are given in Table \ref{tab:cellular system} according to \cite{LTE01}, \cite{Goldsmith01}. Here, users (1,2) employ (QPSK, 8PSK), respectively. The corresponding modulations for IA-based schemes are (4PAM, 8PAM). The number of channel realizations are 1000. As a reference, we also present the SER performance of a point-to-point SISO Rayleigh fading channel with QPSK transmit constellation. The average SER of this point-to-point communication link can be derived analytically as \cite[(6.61)]{Goldsmith01}
\begin{align}
\mathbb{E}\left[ \text{SER}_{\text{QPSK}} \right] \approx \int_{0}^{\infty} 2Q\Bigg( \sqrt{ 2\gamma \text{SNR} }  \Bigg) e^{-\gamma} d\gamma = 1 - \frac{ \sqrt{\text{SNR}} } {\sqrt{1 + \text{SNR} }},
\end{align} 
and, as the SNR $\to\infty$, $\mathbb{E}\left[ \text{SER}_{\text{QPSK}} \right]$ $\to$ $\frac{1}{2 \text{SNR}}$.

\begin{table}[t]
\centering
\caption{\textsc{Cellular System Parameters \cite{LTE01}, \cite{Goldsmith01}}}\label{tab:cellular system}
\begin{tabular}{l l}
\hline
\hline
\vspace{-0.0in} & \\
Cell radius & 1 km \\
Path-loss model & $10\log_{10}(d^{-\mu})$ \\
Path-loss exponent & $\mu = 3.7$ \\
Small-scale fading & $h_{ij}$ $\sim$ $\mathcal{CN}(0,1)$, $i, j \in \{1, 2\}$ or $\{1, 2, 3\}$ \\
Bandwidth & $\Omega$ = $10$ MHz \\
Transmit power & 7 - 57 dBm \\
AWGN power & $\sigma_k^2$ = -174 dBm/Hz 
\vspace{+0.03in}\\
\hline
\end{tabular}
\end{table}

Again, we observe in Fig. \ref{fig:2IC_SER fading} that the SER performance of PS-PC and MSE-based schemes saturates as the SNR increases. In contrast, Minmax-PEP/SER and IA-based schemes achieve  decreasing maximum average SERs over the SNR. It is not surprising, since in the two-user $2\times 2$ SISO IC, the interference signal can be nulled so that the desired signals are interference-free each with diversity order of $1$. The maximum SERs of Minmax-PEP/SER and IA-based schemes are therefore identical and all follow the scaling law $\frac{1}{\text{2SNR}}$ asymptotically as SNR $\rightarrow$ $\infty$ . Certainly, there is a performance gap between the single-user system and two-user SISO IC. Furthermore, the results in Fig. \ref{fig:2IC_SER fading} imply that even for the two-user case, the proposed Minmax-PEP/SER achieves the same performance as the IA-based schemes averaged over random channel conditions, while Fig. \ref{fig:2IC_SERfig1} only shows a particular channel setup where IA-based schemes outperforms the proposed designs.

\begin{figure}[t]
 		\centering 
 		\epsfxsize=0.48\linewidth
 		\captionsetup{width=0.75\textwidth} 
 		\includegraphics[width=14.5cm, height=10.5cm]{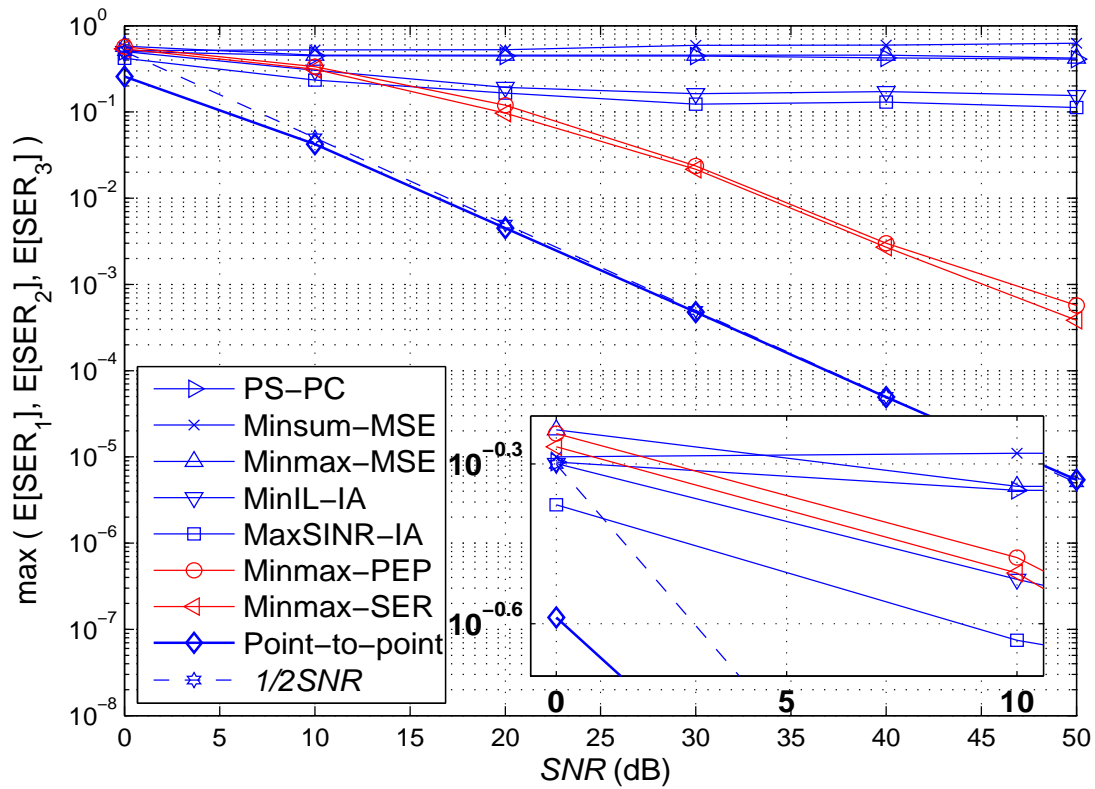}
 		\caption{The SER performance comparison in a cellular network with 3 cell-edge users. }\label{fig:3IC_SER fading}
\end{figure}

In Fig. \ref{fig:3IC_SER fading}, we compare the maximum average SERs of Minmax-PEP/SER with other schemes under a three-cell cellular system (see Fig. \ref{fig:3IC fading}) with parameters also given in Table \ref{tab:cellular system}. The number of channel realizations are again 1000. The users all employ QPSK modulation, and therefore the corresponding modulation for IA-based schemes is 4PAM. From Fig. \ref{fig:3IC_SER fading}, we note that the maximum average SERs of PS-PC, MSE-based, and IA-based schemes all saturate over SNR. In contrast, the minimized maximum SERs of Minmax-PEP and Minmax-SER both decrease over SNR, and also seem to follow the scaling law $\frac{1}{\text{2SNR}}$. However, a detailed analysis, which is out of this paper's scope, is necessary before any conclusion on the diversity order is drawn.    

From the above results in Sections \ref{subsec:awgn} and \ref{subsec:cellular}, we observe that Minmax-PEP/SER are good transmission designs for SISO ICs under both AWGN and cellular system setups. In terms of maximum SER, they guarantee a competitive performance in the two-user case and outperform all other schemes in the $K$-user ($K\geq 3$) case. The advantage of the proposed schemes over PS-PC, MSE-based, and IA-based designs is due to the fact that they fully exploit the CSI and constellation information of the users, represented by the channel coefficients $h_{kl}$'s and matrices $\big\{ \mv{Q}_k \big\}_{k=1}^K$ or $\big\{ \mv{F}_k \big\}_{k=1}^K$, respectively. The Minmax-PEP/SER we proposed therefore provide a significant improvement over existing schemes, which do not take into account the constellation information of the users.

\section{Conclusion}\label{sec:conclusions}

In this paper, we have studied the problem of minimizing SERs for the $K$-user SISO IC with improper signalling applied over practical modulations under fixed data rates. We have proposed two efficient algorithms to jointly optimize the precoding matrices of users to directly minimize the maximum PEP/SER, by exploiting the given signal constellations and the additional degrees of freedom provided by improper signalling. Several benchmark schemes based on conventional proper signalling as well as other state-of-the-art improper signaling designs are compared, while it is shown by simulations that our proposed improper signalling schemes achieve a competitive or improved SER performance in SISO-ICs under both AWGN channels and cellular systems. Our study provides a different view on transmission optimization for ICs in contrast to the large body of existing works on the rate optimization of ICs. 

\appendices

\section{Proof of Lemma \ref{lemma:prob d1-->tilde d1}}\label{proof:prob d1-->tilde d1}

We first derive the PEP $Pr\{\mv{d}_k \to \mv{\tilde{d}}_k|\mv{d}_k\}$ for an arbitrary decoding matrix $\mv{R}_k$ given the Euclidean-distance based detector, and then substitute $\mv{R}_k$ $=$ $\mv{W}_k^{-1/2}$ to obtain Lemma \ref{lemma:prob d1-->tilde d1}. We have
\begin{align}
Pr\{\mv{d}_k \to \mv{\tilde{d}}_k|\mv{d}_k\}
& = Pr \Big\{ ( \mv{r}_k - |h_{kk}| \mv{R}_k^T \mv{A}_k \mv{\tilde{d}}_k )^T
( \mv{r}_k - |h_{kk}| \mv{R}_k^T \mv{A}_k \mv{\tilde{d}}_k ) \notag \\
& \qquad \qquad \qquad \qquad \qquad \qquad \leq  
( \mv{r}_k - |h_{kk}| \mv{R}_k^T \mv{A}_k \mv{d}_k )^T 
( \mv{r}_k - |h_{kk}| \mv{R}_k^T \mv{A}_k \mv{d}_k )  \Big\} \notag \\ 
& = Pr\Big\{  \left[ |h_{kk}| \mv{R}_k^T \mv{A}_k (\mv{d}_k - \mv{\tilde{d}}_k) + \mv{R}_k^T \mv{w}_k  \right]^T 
\left[ |h_{kk}| \mv{R}_k^T \mv{A}_k (\mv{d}_k - \mv{\tilde{d}}_k) + \mv{R}_k^T \mv{w}_k  \right] \notag \\  
& \qquad \qquad \qquad \qquad \qquad \qquad \qquad \qquad \qquad \qquad \qquad \qquad \leq  ( \mv{R}_k^T \mv{w}_k  )^T ( \mv{R}_k^T \mv{w}_k )    \Big\}   \notag \\ 
& = Pr\left\{ |h_{kk}|^2 (\mv{d}_k - \mv{\tilde{d}}_k)^T \mv{A}^T \mv{R}_k \mv{R}_k^T \mv{A} 
(\mv{d}_k - \mv{\tilde{d}}_k) + 2 |h_{kk}| \mv{w}_k^T \mv{R}_k^T \mv{A} 
(\mv{d}_k - \mv{\tilde{d}}_k) \leq 0       \right\}      \notag \\
& = Pr\left\{ |h_{kk}| (\mv{d}_k - \mv{\tilde{d}}_k)^T \mv{A}^T \mv{R}_k \mv{R}_k^T \mv{A} 
(\mv{d}_k - \mv{\tilde{d}}_k) + 
\underbrace{ 2(\mv{d}_k - \mv{\tilde{d}}_k)^T \mv{A}^T \mv{R}_k \mv{w}_k }_{\triangleq D} \leq 0 \right\}. \label{eq:A1}
\end{align}

Note that the elements of $\mv{d}_l$ are i.i.d. $\sim$ $\mathcal{N}(0,1)$, $l=1, \dots, K$, $l\neq k$. Therefore $\mv{w}_k$ is a real Gaussian random vector with zero mean and covariance matrix $\mv{W}_k$ given in (\ref{eq:INR CovMat}), i.e., $\mv{w}_k$ $\sim$ $\mathcal{N}\left( \mv{0}, \mv{W}_k \right)$. As a consequence, the RV $D$ defined in (\ref{eq:A1}) is distributed as $\mathcal{N}\left( 0,4(\mv{d}_k - \mv{\tilde{d}}_k)^T \mv{A}^T \mv{R}_k \mv{R}_k^T \mv{W}_k \mv{R}_k \mv{R}_k^T \mv{A}(\mv{d}_k - \mv{\tilde{d}}_k)  \right)$. From (\ref{eq:A1}), we thus have
\begin{align}
Pr\{\mv{d}_k \to \mv{\tilde{d}}_k|\mv{d}_k\}  = Q\left(  \frac{ |h_{kk}| (\mv{d}_k - \mv{\tilde{d}}_k)^T \mv{A}^T \mv{R}_k \mv{R}_k^T \mv{A} (\mv{d}_k - \mv{\tilde{d}}_k) }   {  2\sqrt{ (\mv{d}_k - \mv{\tilde{d}}_k)^T \mv{A}^T \mv{R}_k \mv{R}_k^T \mv{W}_k \mv{R}_k \mv{R}_k^T \mv{A} (\mv{d}_k - \mv{\tilde{d}}_k) }  }  \right).
\end{align}

Substituting $\mv{R}_k$ $=$ $\mv{W}_k^{-1/2}$, we obtain (\ref{eq:PEP optimal decoder}). This completes our proof of Lemma \ref{lemma:prob d1-->tilde d1}.

\section{Proof of Proposition \ref{prop:P2a P2b}}\label{proof:P2a P2b}

Consider the optimization problem (P1d). The $(i,i)$-th elements of $\mv{G}_k$ are given in (\ref{eq:E1E2_ii}). Note that $\left[ \mv{G}_k \right]_{ii}$ is only dependent on $\mv{b}_{k,i}$, and hence the optimization problem for each $i$ is decoupled with respect to $\mv{b}_{k,i}$. Since the problem is convex for each $\mv{b}_{k,i}$, the minimum point can therefore be obtained by setting the gradient of $\left[ \mv{G}_k \right]_{ii}$ over $\mv{b}_{k,i}$ to zero, i.e.,
\begin{align}
\frac{\partial \left[ \mv{G}_k \right]_{ii}}{\partial \mv{b}_{k,i}} = \sigma_k^2 \mv{b}_{k,i}^T + 2 \sum_{l=1, l\neq k}^K |h_{kl}|^2 \mv{J}(\phi_{kl}) \mv{A}_l \mv{A}_l^T \mv{J}^T(\phi_{kl})  \mv{b}_{k,i}^T - 2 |h_{kk}| \mv{q}_{k,i} \mv{A}_k^T = \mv{0}, ~\forall i=1,\dots,\bar{D}_k.
\end{align}

This gives the following solution
\begin{align}
\mv{b}_{k,i}^T = |h_{kk}| \mv{q}_{k,i} \mv{A}_k^T \left( \frac{\sigma_k^2}{2}\mv{I} + \sum_{l=1, l\neq k}^K |h_{kl}|^2 \mv{J}(\phi_{kl}) \mv{A}_l \mv{A}_l^T \mv{J}^T(\phi_{kl})) \right)^{-1}, ~\forall i=1,\dots,\bar{D}_k.
\end{align}

Therefore, the solution for the matrix $\mv{B}_k$ is
\begin{align}
\mv{B}_{k}^T = |h_{kk}| \mv{Q}_{k} \mv{A}_k^T \left( \frac{\sigma_k^2}{2}\mv{I} + \sum_{l=1, l\neq k}^K |h_{kl}|^2 \mv{J}(\phi_{kl}) \mv{A}_l \mv{A}_l^T \mv{J}^T(\phi_{kl}) \right)^{-1}.
\end{align}

Subsituting $\mv{B}_k$ into (\ref{eq:E1E2}), we obtain
\begin{align}
\mv{G}_k = - |h_{kk}|^2 \mv{Q}_{k} \mv{A}_k^T \left( \frac{\sigma_k^2}{2}\mv{I} + \sum_{l=1, l\neq k}^K |h_{kl}|^2 \mv{J}(\phi_{kl}) \mv{A}_l \mv{A}_l^T \mv{J}^T(\phi_{kl}) \right)^{-1} \mv{A}_k \mv{Q}_{k}^T.
\end{align}

The optimization problem (P1d) is thus equivalent to the original problem (P1c). This completes our proof of Proposition \ref{prop:P2a P2b}.


\begin{thebibliography}{99}
\bibliographystyle{IEEEbib}

\bibitem{Etkin01} R. H. Etkin, D. N. C. Tse, and H. Wang, ``Gaussian interference channel capacity to within one bit,'' \emph{IEEE Trans. Inf. Theory}, vol. 54, no. 12, pp. 5534-5562, Dec. 2008.

\bibitem{Dahrouj01} H. Dahrouj and W. Yu, ``Coordinated beamforming for the multicell multi-antenna wireless systems,'' {\it IEEE Trans. Wireless Commun.}, vol. 9, no. 5, pp. 1748-1795, May 2010.

\bibitem{Zhang01} R. Zhang and S. Cui, “Cooperative interference management with MISO beamforming,” \emph{IEEE Trans. Sig. Proc.}, vol. 58, no. 10, pp. 5450-5458, October 2010.  

\bibitem{Shang02} X. Shang, B. Chen, and H. V. Poor, ``Multiuser MISO interference channels with single-user detection: optimality of beamforming and the achievable rate region,'' {\it IEEE Trans. Inf. Theory}, vol. 57, no. 7, pp. 4255-4273, July 2011.

\bibitem{Bjornson01} E. Bjornson, R. Zakhour, D. Gesbert, and B. Ottersten, ``Cooperative multicell precoding: rate region characterization and distributed strategies with instantaneous and statistical CSI,'' {\it IEEE Trans. Signal Process.}, vol. 58, no. 8, pp. 4298-4310, Aug. 2010.

\bibitem{Jafar01} S. A. Jafar, ``Interference alignment: a new look at signal dimensions in a communication network'', {\it Foundations and Trends in Communications and Information Theory}, vol. 7, no. 1, pp 1-136, 2011.

\bibitem{Schreier01} P. J. Schreier and L. L. Scharf, \emph{Statistical Signal Processing of Complex Valued Data: The Theory of Improper and Non-Circular Signals}, Cambridge University Press, 2010.

\bibitem{Picinbono01} B. Picinbono and P. Chevalier, ``Widely linear estimation with complex data,'' \emph{IEEE Trans. Sig. Process.}, vol. 43, no. 8, pp. 2030-2033, Aug. 1995.

\bibitem{Schreier02} P. J. Schreier, L. L. Scharf, and C. T. Mullis, ``Detection and estimation of improper complex random signals,'' \emph{IEEE Trans. Inf. Theory}, vol. 51, no. 1, pp. 306-312, Jan. 2005.

\bibitem{Navarro-Moreno01} J. Navarro-Moreno, M. D. Estudillo, R. M. Fernandez-Alcala, and J. C. Ruiz-Molina, ``Estimation of improper complex-valued random signals in colored noise by using the Hilbert space theory,'' \emph{IEEE Trans. Inf. Theory}, vol. 55, no. 6, pp. 2859-2867, Jun. 2009. 

\bibitem{Cadambe02} V. Cadambe, S. A. Jafar, and C. Wang, ``Interference alignment with asymmetric complex signaling -- settling the Host-Madsen-Nosratinia conjecture,'' \emph{IEEE Trans. Inf. Theory}, vol. 56, no. 9, pp. 4552-4565, Sep. 2010. 

\bibitem{Ho01} Z. K. M. Ho and E. A. Jorswieck, ``Improper Gaussian signaling on the two-user SISO interference channel,'' \emph{IEEE Trans. Wireless Commun.}, vol. 11, no. 9, pp. 3194-3203, Sep. 2012. 

\bibitem{Zeng01} Y. Zeng, C. M. Yetis, E. Gunawan, Y. L. Guan, and R. Zhang, ``Transmit optimization with improper Gaussian signaling for interference channels,'' \emph{IEEE Trans. Sig. Process.}, vol. 61, no. 11, pp. 2899-2913, June 2013.

\bibitem{Park02} H. Park, S.-H. Park, J.-S. Kim, and I. Lee, ``SINR balancing techniques in coordinated multi-cell downlink systems,'' \emph{IEEE Trans. Wireless Commun.}, vol. 12, no.2, pp.626-635, Feb. 2013. 

\bibitem{Zeng02} Y. Zeng, R. Zhang, E. Gunawan, and Y. L. Guan, ``Optimized transmission with improper Gaussian signaling in the K-user MISO interference channel,'' \emph{IEEE Trans. Wireless Commun.}, vol. 12, no. 12, pp. 6303-6313, Dec. 2013. 

\bibitem{Shen01} H. Shen, B. Li, M. Tao, and X. Wang, ``MSE-based transceiver designs for the MIMO interference channel,'' \emph{IEEE Trans. Wireless Commun.}, vol. 9, no. 11, pp. 3480-3489, Nov. 2010.

\bibitem{Chen01} C.-E. Chen and W.-H. Chung, ``An iterative minmax per-stream MSE transceiver design for MIMO interference channel,'' \emph{IEEE Wireless Commun. Lett.}, vol. 1, no. 3, pp. 229-232, June 2012.

\bibitem{Choi01} J. Choi, ``Interference alignment over lattices for MIMO interference channels,'' \emph{IEEE Commun. Lett.}, vol. 15, no. 4, pp. 374-376, Apr. 2011. 

\bibitem{Cahn01} C. R. Cahn, ``Combined digital phase and amplitude modulation communication systems,'' \emph{IRE Trans. Common. Syst.}, vol. CS-8, pp. 150-155, Sep. 1960.

\bibitem{Foschini01} G. J. Foschini, R. D. Gitlin, and S. B. Weinstein, ``Optimization of two-dimensional signal constellations in the presence of Gaussian noise,'' \emph{IEEE Trans. Commun.}, vol. COM-22, pp. 28-38, Jan. 1974.

\bibitem{Beaulieu01} N. C. Beaulieu, ``An infinite series for the computation of the complementary probability distribution function of a sum of independent random variables and its application to the sum of Rayleigh random variables,'' \emph{IEEE Trans. Commun.}, vol. COM-38, pp. 1463-1474, Sep. 1990.

\bibitem{Craig01} J. W. Craig, ``A new, simple and exact result for calculating the probability of error for two-dimensional signal constellations,'' \emph{in Proc. 1991 IEEE Military Commun. Conf.}, vol. 2, pp. 571-575, 1991.

\bibitem{Proadkis01} J. G. Proakis and M. Salehi, \emph{Digital Communications}, 5th ed., McGraw Hill, Singapore, 2007. 

\bibitem{Hochwald01} B. M. Hochwald, T. L. Marzetta, T. J. Richardson, W. Sweldens, and R. Urbanke, ``Systematic design of unitary space–time constellations,'' \emph{IEEE Trans. Inf. Theory}, vol. 46, pp. 1962-1973, Sept. 2000.

\bibitem{Xin01} Y. Xin, Z.Wang, and G. B. Giannakis, ``Space-time diversity systems based on linear constellation precoding,'' \emph{IEEE Trans. Wireless Commun.}, vol. 2, no. 2, pp. 294-309, Mar. 2003. 

\bibitem{Boyd01} S. Boyd and L. Vandenberghe, \emph{Convex Optimization}, Cambridge University Press, 2004.

\bibitem{Grant01} M. Grant and S. Boyd, CVX: Matlab software for disciplined convex programming, version 2.0 beta, http://cvxr.com/cvx, Mar. 2014.

\bibitem{Gomadam01} K. Gomadam, V. R. Cadambe, and S. A. Jafar, ``A distributed numerical approach to interference alignment and applications to wireless interference networks,'' \emph{IEEE Trans. Inform. Theory}, vol. 57, no. 6, pp. 3309-3322, June 2011.

\bibitem{Peters01} S. W. Peters and R. W. Heath, Jr., ``Cooperative algorithms for MIMO interference channels,'' \emph{IEEE Trans. Veh. Tech.}, pp. 206-218, vol. 60, no. 1, Jan. 2011.

\bibitem{LTE01} ``LTE; E-UTRA; RF requirements for LTE pico node B,'' ETSI, Tech. Rep. 136 931 V9.0.0, 2011. [Online]. Available at http://www.etsi.org/deliver/

\bibitem{Goldsmith01} A. Goldsmith, \emph{Wireless Communications}, Cambridge University Press, 2005.

\end{thebibliography}
\end{document}